\documentclass[12pt,a4paper]{elsarticle}
\usepackage[latin2]{inputenc}
\usepackage{graphicx}
\usepackage{ulem}
\usepackage{amsmath}
\usepackage[hyphens,spaces,obeyspaces]{url}
\usepackage{comment}
\usepackage{color}
\usepackage{soul}
\begin{document}
\begin{frontmatter}
\title{Going Viral: The Epidemiological Strategy of Referral Marketing
}
\author[add1]{Sayantari Ghosh}
\ead{sayantari@gmail.com}
\author[add2]{Saumik Bhattacharya}
\ead{soumiksweb@gmail.com}
\author[add3]{Kumar Gaurav}
\ead{gauravk@iitk.ac.in}
\author[add3]{Yatindra Nath Singh}
\ead{ynsingh@iitk.ac.in}
\address[add1]{Department of Physics, Rammohan College, West Bengal 700009, India}
\address[add2]{Indian Statistical Institute, West Bengal 700108, India}
\address[add3]{Indian Institute of Technology Kanpur, Uttar Pradesh 208016, India}
\begin{abstract}
By now, Internet word-of-mouth marketing has established its importance for almost every kind of products. When word-of-mouth spread of a marketing campaign goes viral, the success of its strategy has been proved to be path-breaking for the firms. The current study is designed to extend knowledge about customer response to these campaigns, and is focused on the motivations that lead them to different responses. Primary goal of this study is to investigate the reasons that drive this dynamics and to generate a justified theoretical framework of diffusion dynamics of viral marketing campaigns.\\
A rigorous analysis of data obtained through a questionnaire-based survey helped us to understand how people, who are from different age group, sex and locations, define and evaluate the encounter with a referral campaign online, in similar and unique logical ways. The key finding was a conceptual framework of customer motivation which in turn dictates the dynamics of the campaign. We have identified important themes related to customer motives, like significance of: incentives and ease,  inherent forgetting, reminders from peers compared to company's remarketing mails, trust and brand value etc.\\
Drawing an analogy with differential equation-based models of infectious disease spread, this paper further provides some initial evidence that participation in viral marketing campaigns has several consumer related dynamical factors which can be incorporated in an epidemiological model for mathematical treatment, to indicate key operational factors in ensuring an effective spread of the marketing campaign.
\end{abstract}
\end{frontmatter}
\section{Introduction }

The field of marketing has witnessed a revolution, as the customers have become more urbane and focused during the last decade, which has caused the introduction of new marketing technologies through the Internet.  E-commerce, social networking websites and online life brought noticeable growth in digital marketing. According to World Internet Users Statistics, 51\% of the world population have found their way online, and this number is rising everyday \cite{WIUS, o2004electronic}. While the advertising formats such as the press, television and outdoor media still have their own perks, Internet word-of-mouth marketing, which is also popularly known as Viral Marketing (VM), is emerging as the corner stone marketing strategy in recent times. In VM campaigns, the consumers themselves are encouraged to share their product preference through their social networks, generating brand awareness among a large population extremely fast. \cite{meerman2008}. As customers nowadays bank on the online social media to find the truth, they wish to validate claims about a product by a genuine personal response. According to traditional definition and understanding, the principals of VM rests on exploiting pre-existing social networks (or other online platforms, like, web forums,  blogs, e-mails etc.), to accomplish precise marketing goals \cite{cruz2008evaluating}, by considering the existing customers as brand advocates. The domain of viral marketing has grown vastly over past decade, which now includes passing along advertisements,  photos, videos, promotional hyperlinks, animations, games, newsletters, press releases, or whatever else that can advertise a particular brand.\\
In 2003, Mohammed et al.  \cite{mohammed2003internet} pointed out that users are less likely to trust promotional communications by the company themselves compared to the peer recommendations about a product or service.  In 2012 Nielsen's Global Trust Advertising survey \cite{Chaney2012} reported, ``\textit{people don't trust  advertising, at least not as much as they trust recommendations from friends and consumer opinions expressed online."} With a data for more than 28,000 Internet respondents in 56 countries, it has been seen that 92\% of consumers trust recommendations from friends and family above all other forms of advertising. This is one of the major reasons why products that use a VM campaign tend to succeed very quickly. Moreover, it is rather clear that VM would cost considerably less than the traditional marketing and promotional techniques.  Starting from gaining new patrons \cite{godes2009firm} to creating brand intimacy in existing customers  \cite{blazevic2013beyond}, this strategy of Internet word-of-mouth creates several positive outcomes for the firm. Customers proceed with brand recommendations for diverse reasons: while someone joining in a referral program may endorse a brand to earn a monetary incentive \cite{ryu2007penny}, another may do the same to communicate identification with the brand \cite{berger2012makes,wojnicki2008word}.\\
Turning the Visitors into ambassadors several booming growths have been seen in recent times. While there is no shortage of news publications fighting for the attention of millennial, in 2017, the San Francisco-based daily email newsletter the Hustle gained 300,000 subscribers in just a few months, with great copywriting and an aptly planned Milestone Referral program. With a belief that the most powerful and effective way to launch their grooming brand was through a credible referral, in 2013, Harry's, a New York-based shaving equipment manufactures gathered 100,000 customers in one week before they launched. They have used a strategically planned Milestone Referral campaign where the minimum required referral numbers were kept as low as 5, and 77\% of their initial customers were collected via referral, where 20,000 people referred about 65,000 friends. Dropbox's referral program is possibly one of the most famous cases of viral marketing
executed with exceptional success. Dropbox's metric history shows a 3900\% growth in within 15 months (September 2008: 100,000 registered users to December 2009: 4M registered users), with a 500MB for 1 referral marketing idea, coupled with an easy invitation process and clear view of the benefits. In another extraordinarily successful strategy with a well-timed and well-implemented referral marketing program, Airbnb offered \$25 discount for accommodation booking to both sides, which became popular as the `altruistic referral'. Rodrigues et al. \cite{rodrigues2016can} first pointed out that as soon as a referral online marketing message goes viral, it draws a close analogy to an epidemic, involving transmission through contacts and spread in a population. They have proposed a mathematical model of the VM progression, using insights from epidemiology.
\section{Rationale for the study}
Although prior research has added to the current knowledge of VM dynamics, studies that attempt to balance qualitative perspectives with quantitative methods are lacking. While attempting mathematical treatment of the problem is an interesting way to address the issue, studies are essential to identify the impact of various external factors related to the consumers that actually influence the propagation of viral campaigns. The aims of this study are twofold:
\begin{itemize}
\item First, here we use a combination of quantitative and qualitative methodology to investigate the motives reported by the consumers for sharing or not sharing a viral campaign. The whole study is rooted in data collected through an extensive questionnaire-based-survey where we take inputs from customers from different age groups belonging from different parts of India.
\item Next, we develop a conceptual framework of VM participation (or withdrawal) motivation. We compare our findings from the survey with a standard epidemiological model, known as SIR (susceptible-infected-recovered) model, and try to point out the major similarities and differences.
\end{itemize}
This is an attempt to clarify the ideas about the flow of a promotional advertisement with the means of referrals, so that a data-driven understanding of the actual dynamics of VM campaigns can be developed.
\section{Methodology}
The main agenda for this study was concerned with the design, circulation and interpretation of a questionnaire, to find out how the participants have
experienced of viral marketing, and their idea, positive or negative, towards these
marketing schemes. The questions were supposed to be designed in a way so that they remain accessible and easily understandable to the participants, while shedding light on the key factors that affect the customers to share or not to share a viral
campaign. The participants were supposed to be chosen keeping it as diverse as possible from the perspective of gender, age and background.
\subsection{Questionnaire}
The questionnaire has been designed to identify the dynamics of viral marketing in a real-world environment (see Appendix \ref{sec:appA}). It was decided that the questionnaires would be shared with the prospective participants via e-mails, WhatsApp, Facebook and printed forms to increase the number of contributors, so that the results can be better generalized. No constraint has been imposed on the participants to complete the questionnaire within a short time, to avoid  \cite{brace2018questionnaire} unnecessary anxiety or stress and allows the participant to respond in their own time. In addition, the participant anonymity was ensured to encourage truthful answers and better results \cite{brace2018questionnaire}.
\subsubsection{Questionnaire Design }
The questionnaire was designed keeping a balance, so that the answers can supply us with the inputs that we needed from the participants, though answering it was kept as simple as possible. We designed a questionnaire which had a logical structure and was split into three defined parts (see Appendix \ref{sec:appA}). To ensure participants are clear on the information they are required to provide, we explained the interest and idea behind this study through a small write-up before the questionnaire began. The first part of the questionnaire clarified whether a participant had been involved in viral email marketing campaigns ever, while the other two segments were to explore the reasons that caused his/her interest to grow or decay about a certain campaign. All questions were sequentially related to point out some overall important factors that drive this dynamics.
\\
The questionnaire included both open questions, in which flexible answers were accepted from the participants, and closed questions, for which choosing from given options was mandatory. Out of the total of 18 questions, \textbf{ten} questions asked for a  polar (i.e., yes/no) response; results obtained as the outcome of these questions were absolutely clear and defined. This also guarantees that with only two options to choose from, the participants will take a careful decision. This method has been found to enhance precision as the answers categories could be easily interpreted \cite{leung2001design}. However as these questions narrow the scope of responses, the questionnaire also contained \textbf{five} questions that had a room for the participant to give a detailed and qualitative answer, or to include an explanation for particular choices they made. This in turn provides the researcher with valuable information to get a wider view of it and keeps the research reliable and valid \cite{leung2001design}. However, these questions also create the chances of a non-response occurring for those participants who try to avoid longer answers. With both types of questions in it, our survey seemed to address the issues associated and to enhance response rate as much as possible.
\subsubsection{Participants}
This questionnaire was sent to more than seven hundred prospective participants from all around India. While 62\% of them were contacted through email, 23\% were sent requests through popular phone applications like WhatsApp, Facebook and Twitter. 15\% of the total sample were directly approached with a printed version of the questionnaire. At least one relevant contact information associated with the prospective participant (i.e., email IDs, contact numbers, Facebook page link, name \& address etc.) was recorded in an address book. All the participants are residents of India, but they belong from different parts of the country; their age ranged between eighteen and seventy. For initial contact, equal numbers of males and female participants were chosen. A helpline number as well as an email-id was provided for clarification of any doubts regarding the questionnaire.
\subsubsection{Data Collection Method }
In case of the online distribution of the questionnaire (via email, WhatsApp, Facebook etc.), we have sent a link to the survey constructed through Google Form. The message was preceded by a cover letter (see Appendix \ref{sec:appB}) to emphasize on the goal of the study and to convince the participants to take part in the survey \cite{leung2001design}. In case of questionnaire handouts, the printed versions were distributed among the participants and drop-boxes were placed at informed locations, so that the participants can submit their responses in their own time and keeping their anonymity intact.  Each completed questionnaire, collected via Google form, or by hard-copy submission, was archived, with a time-stamp as well as a serial number, in an Excel file.
\subsubsection{Ethics}
The purpose of the cover letter was twofold: along with explaining the purpose and importance of the survey, it also ensured that the respondents were not deceived in any way.  The cover letter clearly notified the participants that by responding through the link associated with the email they were giving their informed consent (see Appendix \ref{sec:appB}). The confidentiality and anonymity of the data provided was assured from the researchers' end through the same mail. As a matter of fact, the questionnaire asked for no personal information, like name or location of the participants.

\section{Results and Observations}
The data obtained through the questionnaire resulted into some very interesting observations
which were analyzed to understand the true propagation of viral marketing. Of total 715 sampled email addresses, 331 people were successfully recruited to join as participants, with a response rate of 46.29 \%. The gender breakdown of the respondents was 65.7\% male and 34.3\% female. 57\% participants were of young age (18-30), 35\% were middle aged (31-55) and 8\%
senior (56-70). The qualitative analysis of the survey data generated a conceptual framework of consumer motivation to actively participate in a viral campaign, which we discuss in detail in Sec. \ref{framework}. This framework points out that whether to be active or to remain dormant depends on how every individual define and evaluate the campaign, the social network around them, and their own agency at that point in time. We also identify a set of activity domains emerging as an outcome of the survey, that are found to steer the flow of the campaign with major effects: (a) reward and referrals, (b) spontaneous forgetting, (c) company remarketing vs. friends reminding, (d) trust and security, coupled with the idea of brand. Let us elaborate on these observations in the following sections.
\begin{figure}[ht]
\centering
\includegraphics[width=12cm]{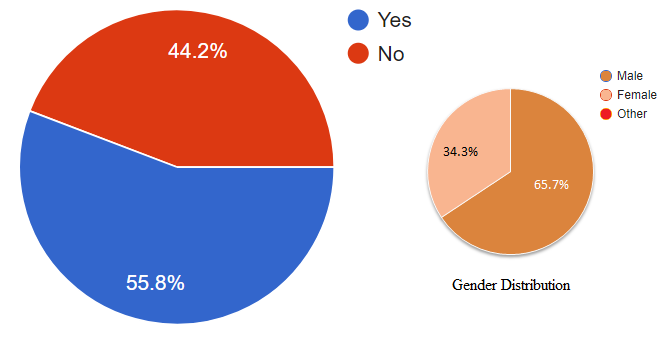}
\caption{ Interest to share/take part in a VM campaign. The question asked was polar in nature and the answer estimated what percentage of the total participants (of which 65.7\% were male and 34.3\% were female: Inset) shared an online advertisement to avail an offer/ reward/discount.}
\end{figure}
\subsection{Rewards and Referrals}\label{ssce:reward}
The incentive or reward, provided by the advertiser to the advocating customer, is found to influence their motivation of participating in a referral program a lot. In our survey 55.8\% of the total participants admitted that they shared an online advertisement with a friend (or shared the contact detail of the friend) to avail an offer/reward/discount. Among these, for 74\% one of the major reasons for sharing was how lucrative the offer was for themselves. It is evident that amount of reward is a prominent driving force behind the VM marketing. Out of 154 participants, 55.8\% said that they will definitely be interested in an offer that someone decided to ignore, if he/she hears that his/her friends have won a good reward from it and 37.7\% participants reported that they may be interested in the offer in the same scenario. Though, reward is one of the key aspects of VM, there are other factors that directly control a VM spread. An interesting observation that came out from the survey is that even if a VM offer is genuine,  62\% participant who were actively sharing an offer got bored and lost interest in sharing the message. From this a contradictory conclusion can be drawn: through incentives, the customers feel more motivated to recommend, but these referrals may not appear as genuine to the receivers, as soon as they get to know the recommendations are driven by incentives, and not by customers' personal experience with the product. This apparent problem can be solved using strategically-planned two-way incentives, where the customer who referred someone will earn a credit, discount or money, and the referral will usually earn a discount. Two-way rewards lead to many successful referrals as both parties have a reason to engage in the program. AirBnB referral marketing demonstrated an effective implementation of this; Uber's two-sided referral system was also very successful.\\
Along with the incentives, another driving force of sharing motivation was ease of the sharing procedure and flexibility of the number of referrals. A considerable percentage of  customers claimed that they only proceed for sharing if the process is not too complex (40.9\% of total population) and they will not be forced to share in bulk (58.8\% of total population) (Question no. 1, Module 2, Options: ``\,Minimum number of sharing required" and ``\,Easy to share"). If we look back on some successful referral marketing campaigns that received a viral response we can see that, while for Dropbox or AirBnb the customers can select any number of referrals they want (with a minimum of only \textit{one}), for the Hustle and Harry's, the first milestone was set at four and five recommendations respectively. All these numbers, being quite low in value, are in accordance with our study results. From these discussions it can be concluded that the approximate ratio of the number of referrals and the amount of incentive acts as a strong factor that drives the sharing dynamics.
\begin{figure}[ht]
\centering
\includegraphics[width=12cm]{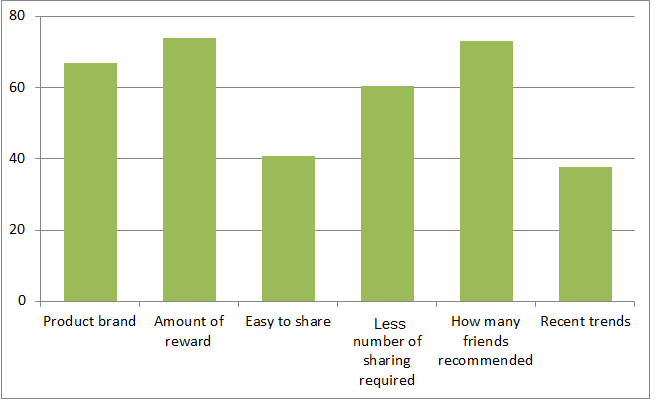}
\caption{Reasons behind sharing an online advertisement. Y-axis indicates the percentage of people motivated by the reasons mentioned along X-axis. Participants could choose more than one option according to their choice.}
\label{fig:reason_sharing}
\end{figure}
\subsection{Spontaneous Forgetting \& Attention Diversion}
A key finding of our survey was related to the inherent forgetting associated with the customers. The concerned question was Question no.2 of Module 2 (Appendix \ref{sec:appA}) which was only asked to those who expressed their interest as well as responded as being active part of a VM progression before. The question asked the participants whether they ever missed a viral offer, which they first decided to avail. In answer to our question, 76.6\% participants answered affirmatively. Focusing on the reasons that made them miss the offer, diversion of attention and forgetting were detected as the two major contributors (59.1\%, 50.4\%).\\
Internet is flooding with information nowadays, but sources for persistent information are rare in social networking websites. In social media, people shift their focus myopically on the information of the moment without spending much the time to form meaningful opinions. With almost 85 million media uploads per day in popular social networking website like Instagram \cite{brandwatch}, sometimes many interesting campaigns fail to break through the noise and proactively engage their customers. Unfortunately, it creates an unavoidable pool of forgetful or distracted consumers who might have been interested in the product and also could have shared it to spread it among their neighbours. These customers, as previously interested in the advertisement, might be brought back and recruited as the brand advocates if properly identified, or reminded.
\begin{figure}[th]
\centering
\includegraphics[width=12cm]{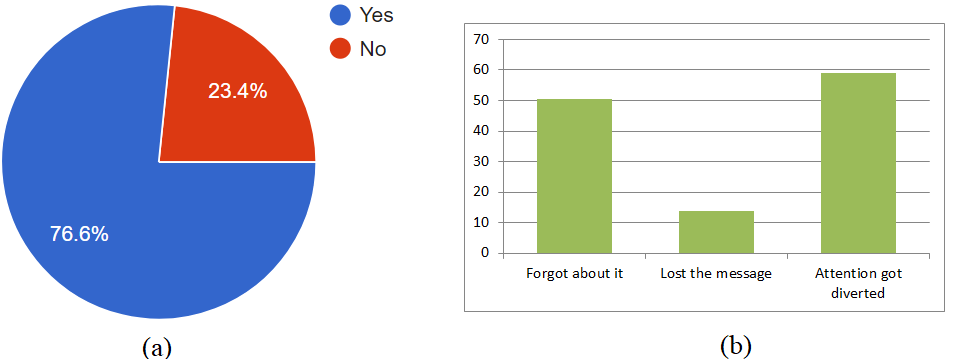}
\caption{(a) Results about missing a viral offer, despite of having initial interest. (a) `Yes' (`No') signifies the person missed (never missed) such an offer. (b) Reasons for missing an offer. Y-axis indicates the percentage of people motivated by the reasons mentioned along X-axis. While forgetting and diversion of attention were the reasons for majority, some people also claimed that they lost the message. Participants could choose more than one option according to their choice.}
\label{fig:missed_offer}
\end{figure}
\subsection{Company's Remarketing vs. Friends' Reminder} \label{ssec:remind}
\begin{figure}[ht]
\centering
\includegraphics[width=12cm]{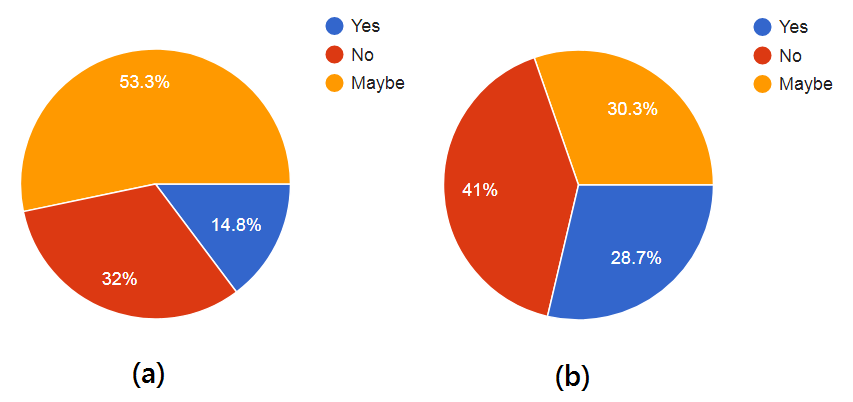}
\caption{Response from the people initially showing no interest towards viral marketing campaigns, about (a) gaining interest if they read positive reviews of the product/offer in company website or company forwarded promotional email, and (b) becoming interested in an campaign/brand if it is frequently mentioned by friends or in their social circle. `Yes'(`No') signifies people who (do not)  think the reason will motivate them; ``May be" signifies they are unsure about it. }
\label{fig:comapny_reminder}
\end{figure}

Remarketing is a method of advertising that serves ads to the people who have already expressed interest in a product or a promotional offer, but left the website without completing the desired activity \cite{remarketing}. Nowadays, companies try exceptionally hard to get these customers back using Customer relationship management (CRM) software. To catch the attention of the distracted customers extensive email marketing is used. The most known techniques include exclusive deals or rewards for loyal customers, time limited discount coupons, creation of unique content for newsletter, sneak peeks of future products or events, reminders about abandoned shopping carts etc. to entice them to propagate the message in their own social network. These techniques occasionally succeed in winning back lost, forgetful or distracted customers as brand advocates, but sometimes these mails cause unnecessary annoyance and irritation to the customers. In our survey, as we asked the participants about their response towards company-sent promotional emails, we had less than 15\% customers who were positively enthusiastic about those [see Fig.\ref{fig:comapny_reminder}]; this question also included the positive reviews on company website. Most of the participants seemed unsure (53.3\% ticked ``May be") about their reaction towards these mails; some also mentioned in their detailed responses they mostly don't go through these emails as they receive \textit{``too many such mails"}.
\\
Another factor that came out as an interesting observation of our survey is that reminders by friends, or family members about a viral offer, worked positively for most of our forgetful respondents $[$Question no. 3, Module 2; Appendix \ref{sec:appA}$]$. It is evident that while looking for a
recommendation about a service or a deal, people usually turn to peers and friends to get to know their personal experiences. This also means that the peers have an added access (or opportunity) to remind them about the offers they forgot about. This observation from the survey is supported by the extremely successful VM campaign by Dropbox, where the users knew their referral status by an easily accessible panel through their profiles. In most of online referral campaigns, after the participants complete the steps and invite their friends, no notifications or emails are sent to inform how many of the receivers actually  registered successfully from the referral link. Dropbox included this so that users can see how their invites performed; it opens up a way for friendly reminders, if the user choses to proceed with it.\\
While several participants mentioned the friendly reminder perspective, one of them clearly mentioned ``\,\textit{Reminded by a friend about a(n) offer in a travel app while booking a flight}". This points out another important aspect: as the nearest neighbors of a customer on his
social network (those who are close to them and know them personally), have a close proximity and knowledge about them. Reminders by these people at appropriate time may have \textit{much more probable} effect.\\
Friend recommendations are at the heart of viral marketing; we observed that almost 76\% customers usually go for a product if many friends recommend it. But this observation is beyond that. This influence by peers that successfully makes a person \textit{switch from inert to active state} is often ignored while considering VM dynamics. While the email marketing efforts by the companies have their own perks, all of these strategies have something in common, i.e., automation. Software imposed personalization often ends up annoying the customers. Reminders by friends or family on the other hand is a much more spontaneous, more likely to be aptly timed, and that is why, has more possibility to cause a successful switching.
\begin{figure}[th]
\centering
\includegraphics[width=12cm
]{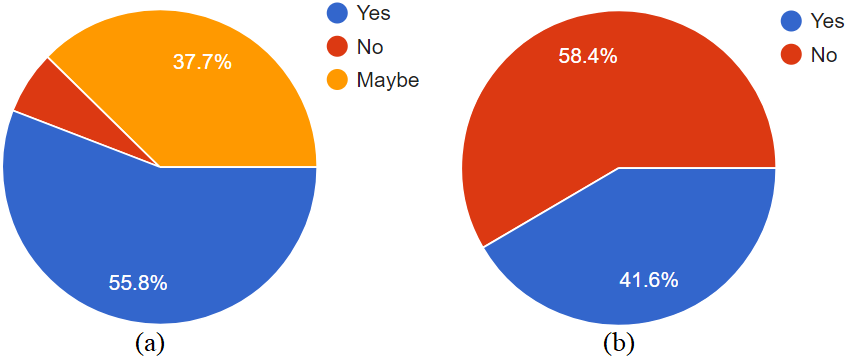}
\caption{Regaining interest in a viral offer  which was ignored or missed when (a) a friend wins a good reward from it and (b) when a friend reminds you about it. `Yes'(`No') signifies people who (do not)  think the reason will cause them regain of interest in an ongoing offer. ``May be" signifies they might regain their interest about it. }
\label{fig:regain_inert}
\end{figure}
\subsection{Trust \& Security: The Brand Value}\label{ssec:trust}
From the perspective of the companies, branding is the tactic aimed to attract people towards their product by creating a method for quick identification, and to provide them with a reason to pick up a product over the competitor \cite{mark2001hero}. Brand is a set of associations and sensations perceived by the customers or potential customers, and they attach value such as trust, identity and purpose, to the product or service offered. A successful brand works with consistency and coherence, balancing the concepts, in all digital touch points to their audience, in social networking, e-marketing, email, landing pages, online services, etc. With the intention of creating proximity to the target audience, companies use this tool to strengthen their identity and authenticity.\\
While the literature has explored many aspects of the impact \cite{chevalier2006effect} and social networking dimensions \cite{goldenberg2006role, katona2011network} of viral marketing, understanding of brand characteristics as precursors to VM is surprisingly limited. In a key observation, we found that not only the product or associated incentives, but the brand plays an important role in the viral dynamics. Among 154 people who took part actively in a viral campaign, 66.9\% admitted that product brand is one of the most important reasons for them. In their detailed answers, we found people associating the product brand directly with the trustworthiness of the campaign (``\,\textit{If\ldots it is legit and i find the concerned corporation trustworthy}"). In a flood of promotional mails and offers, brand provides the customers with an ease to proceed with their interest, not being afraid of security issues. In our survey, one participant explained clearly that while availing an offer, he/she first looks for a way which can ``\,\textit{prove that its not fraudulent activity and only intent is marketing}"; brands provide this peace of mind and save decision-making time for the customers.\\
\begin{figure}[th]
\centering
\includegraphics[width=13cm]{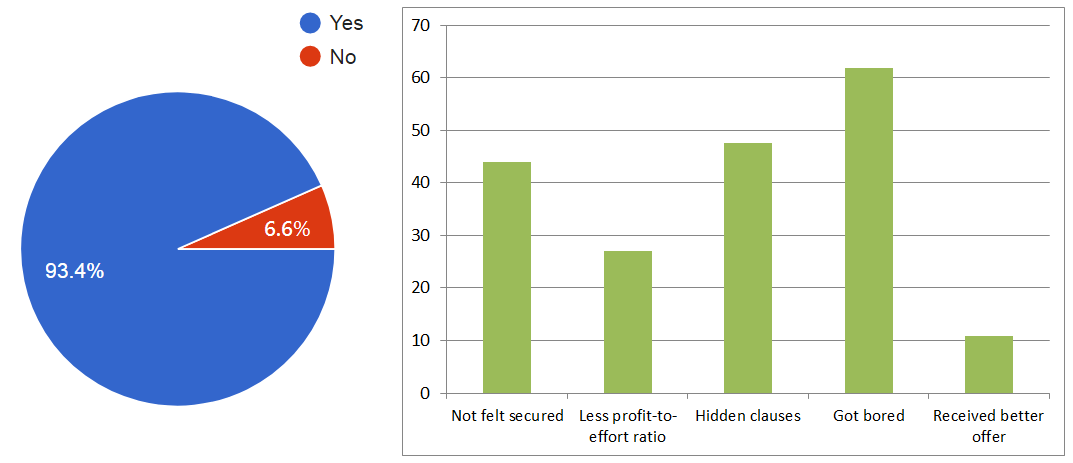}
\caption{Losing interest in an ongoing viral offer: (a) `Yes'(`No') signifies the person (never) lost interest in a campaign. (b) Reasons for losing interest in an offer. Y-axis indicates the percentage of people driven by the reasons mentioned along X-axis.}
\label{fig:lost_interest}
\end{figure}
The observations also point in another direction of dedicated brand advocates. Different scenarios can give rise to some clusters of people, who are especially attached to a particular brand. For example, consumers often prefer brands that are either apparently trendy, or that helps them to fit into a certain group or social circle \cite{chan2012identifiable}. In our survey, 37.7\% respondents agreed that recent trends were an important reason for them while sharing a campaign [see Fig.\ref{fig:reason_sharing}]. When asked precisely about how their decision about a brand is influenced by frequent mention of the related campaign in their own social circle, a major 69\% customers (28.7\% `Yes'. 30.3\% `May be') admitted that they are definitely, or possibly, affected by that [see Fig. \ref{fig:comapny_reminder}] .  The origin of this simply comes from a desire to fit in their respective social circles. In a completely different context, some consumers become habituated, or loyal, to brands that deliver a constant, high-grade performance. Loyalty is essentially affection to a brand, where customers show a strong affinity, and associate the brand with their own personality. Apple's ``\,I'm a PC/I'm a Mac" campaign is a perfect example to demonstrate how brands can get connected with self-perceptions of the users. Whatever may be the origin of these commitments towards a brand, these associations sometimes make people to get interested in only those campaigns connected with their preferred brands; though this is essentially a very local effect, this affinities can play their part in the diffusion dynamics through social networks.\\
If we focus on response of the customers who avoid viral campaigns in general, interestingly, there also we find security to be a major concern. While 48.3\% of the total 120 respondents said that they felt insecure, 47.5\% agreed that a fear for hidden clauses was the reason for
not-sharing for them. As inherently people prefer to avoid risk, VM campaigns associated with unknown brands are generally ignored due to a fear of the unknown, i.e., the safety issues. A well-known established brand name offers safety and reduces the risk of disappointment. People associate familiar brand names with stability and reliability of the product, considering this to be a reason behind the evolution of the brand, and people even pay higher prices for these brands while avoiding unfamiliar brands, because they know what to expect.\\
Another important observation that emerged from the survey is that the people who are not sharing a viral message are concerned with the authenticity with the campaign. Either due to negative past experiences with other spurious campaigns or receiving discouraging word-of-mouth, some participants (12\% out of 122 people who do not participate in VM campaign) are  skeptic about the authenticity and reported that proof of the reward is necessary to convince them that the campaign is not fake. A close observation reveals that such people rely on their trusted friends and family members as evaluators. If a close friend or relative gains an advantage from a VM campaign and share the message, the skeptic inert class becomes convinced and become willing to participate in the campaign. Reminders and communications from the company on the other hand do not have any substantial effect on their decision making as they are unclear about the intention of the company.
\section{Proposed Conceptual Framework} \label{framework}
\begin{figure}[tbp]
\centering
\includegraphics[width=14cm]{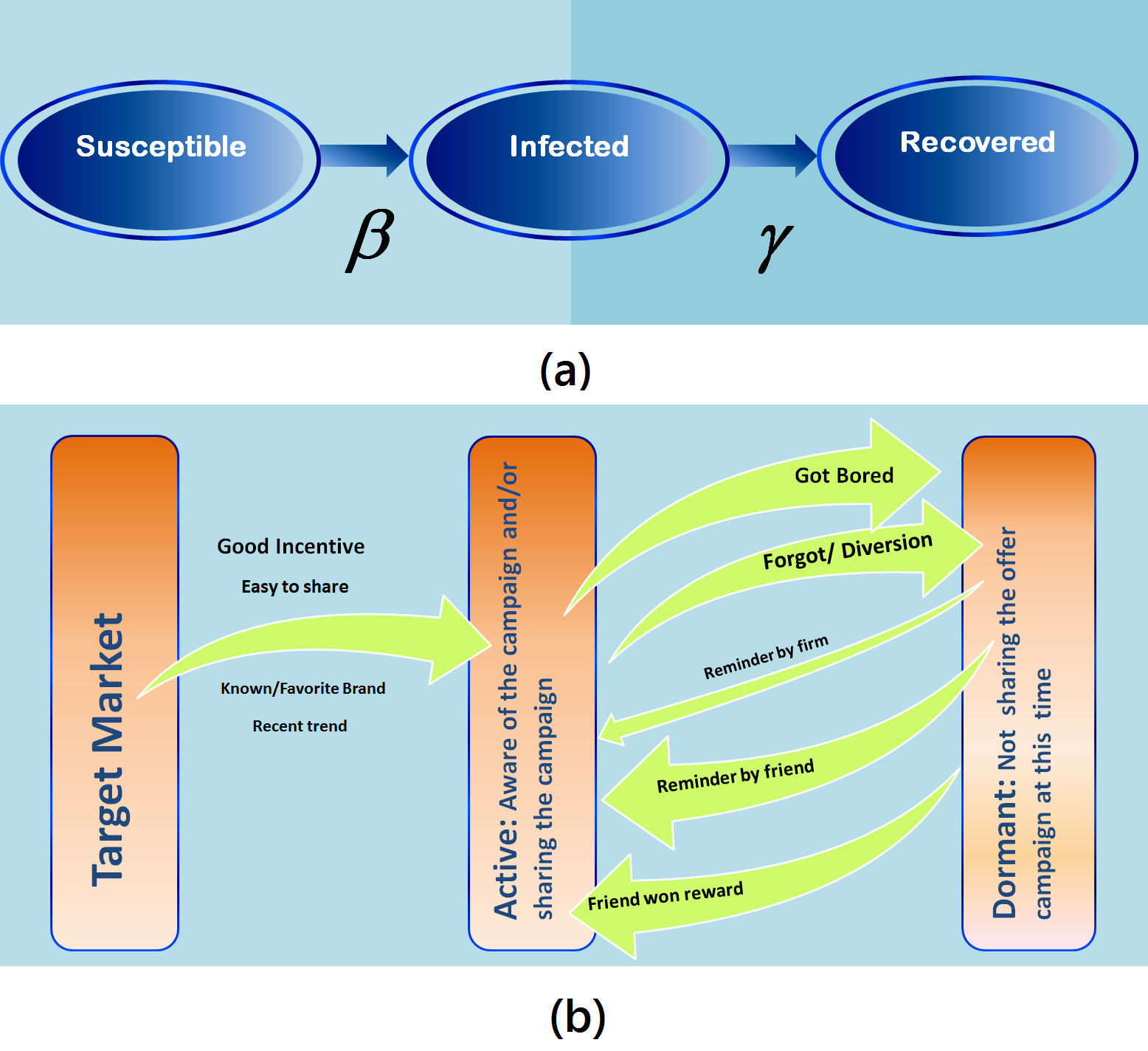}
\caption{Flow of viral campaigns in a population consisting of different community compartments and switching of population behavior depending on different factors: (a) A framework similar to the commonly known S-I-R model of mathematical epidemiology.(b) A more nonlinear population switching framework proposed in our study, entirely based of ground data collected through the survey. Susceptible, Infected and Recovered classes from (a) are equivalent to Target market, Active and Dormant compartments of (b).}
\label{fig:model_dynamics}
\end{figure}

Guided by the data accumulated through the rigorous survey, we now take a logical approach to understand diffusion process of VM campaigns in social networks. Our analyses clearly reveals that beyond the challenge of creative design features, a profound understanding of customer behavior is essential for a firm for managing matters related to relevance and survival of a campaign. Driven by the survey data we attempt to justify, and enrich the epidemiological dynamical model proposed by Rodrigues et al. \cite{rodrigues2016can} to justify the transitions from one class to another. In this work, the authors proposed that the propagation of a VM campaign can be described mathematically using an Ordinary differential equation-based (ODE) epidemic model \cite{leskovec2007dynamics, richardson2002mining, sohn2013viral} of infectious disease spreading. A block diagram representing the assumed framework by them is shown in Fig. \ref{fig:model_dynamics}(a). In this approach, the whole population was assumed to be divided in mutually-exclusive compartments, namely susceptible ($S$), infected ($I$) and recovered ($R$). While $S$ class was indicating the target population, $I$ and $R$ classes indicated the class of people who are potential broadcasters of the message and who are done sharing the message respectively. The model allows switching from one compartment to the other with definite probabilities for moving from the target audience to infective, and from infective to the recovered class; the related parameters are chosen as $\beta$ and $\gamma$. Considering the total population ($N$) to be constant, the model provides a simple and intuitive approach in understanding the VM process through epidemiological perspective.
\\
With this outlook, we attempt to enhance the physical significance of the associated parameters, considering the implications of our survey observations. According to our understanding of this diffusion dynamics, we can actually interpret $\beta$, the effective transition rate from $S$ to $I$ class, as the product of the average number of persuasive contacts per unit time between a target and a broadcaster through their social networks, and the probability of taking part in a campaign after a contact. While the number of contacts will be dictated by the structure of the social network itself, the second contributing factor might depend on several aspects. As we reported in Sec. \ref{ssce:reward}, rewards and benefits, coupled with the minimum no. of referrals needed for participating, came out to be one of the major guiding factors to affect this probability. Ease of share was also pointed out as a vital feature in this regard [see Fig. \ref{fig:reason_sharing}]. Companies should consider these issues while going for a VM campaign. As observed and explained in Sec. \ref{ssec:trust}, familiar brands make people correlate several elements like loyalty, reliability, authenticity etc. with a campaign, and brand names could influence the decision to a considerable extent.\\
Similarly, factors influencing switch from $I$ to $R$ class were taken as forgetting, loss of interest or divergent attentions, whose effects can be accumulated into the rate constant $\gamma$. All these assumptions of  \cite{rodrigues2016can} are quite justified according to the findings of our study, but we found some additional layers to it. We clearly note that there are two specific kinds of people who transit from $I$ to $R$ class; the first group were those who were actively involved and interested about a campaign, and they have participated in broadcasting for the brand for a certain time. These people, as we observe from our survey results, eventually move to the $R$ class mostly because of boredom, low effort-to-gain ratio or due to suspicion of insecurity and hidden clauses [see Fig.\ref{fig:lost_interest}]. On the other hand there is another completely distinct group, who moved to the $R$ class soon after they encounter the advertisement campaign, though they had an initial interest about the offer. The driving causes for these people are usually forgetting, diversion of attention, or simply losing the message [see Fig. \ref{fig:missed_offer}]. All these reasons actually constitute the effective transition rate, $\gamma$.\\
In this context, we must mention that though some of the switching between compartments observed through our survey, supports the transitions of this dynamical model of Fig. \ref{fig:model_dynamics}(a), there exists some important observations (which are connected to some transitions) that cannot be explained using a simple SIR model. To elaborate on that, we must note that the recovered class, $R$, or we can call this class \textit{dormant} for the relevance to the problem, is made of people who are not sharing the message currently, though they are aware of the campaign. We have explained earlier that, though, broadly speaking, these people belong to the same class, the reasons which brought them to the dormant class can be entirely different. People who feel that they missed an offer [see Fig.  \ref{fig:missed_offer}] due to forgetting or by getting caught up in something else (attention diversion), remain to be passively interested about the campaign. If they are reminded by a friend [see Fig. \ref{fig:regain_inert}(b)], or by company emails [see Fig. \ref{fig:comapny_reminder}(a)], as explained in Sec. \ref{ssec:remind}, they immediately gain their active state back, and start contributing in the propagation of the campaign again. Similarly, if we turn our attention to the people who backed up due to security concerns, getting an update about the authenticity from a friend or family member can bring them back to the active `infective' state. In fact, we can also consider the people who got bored of an campaign, or simply ignored them in the first place; a huge percentage among them agree that if an acquaintance wins a considerable reward from the same campaign, they will return to the active participation almost surely [see Fig. \ref{fig:regain_inert}(a)]. It is evident from the above discussion that a simple one-way transition from $I$ to $R$ cannot justify these findings.\\
We propose a conceptual framework of the whole dynamics as depicted by our survey and represent it through a block diagram in Fig. \ref{fig:model_dynamics}(b). The framework is entirely based on the discussions of this section, with inputs from figures and data from the survey. The arrows, as intuitive as it can be, represents transition, while their thickness (or weight) roughly indicates the comparative contribution of a cause. We propose that while several reasons decide the propensity of transition from target (or susceptible, $S$) to active (or infective, $I$) state, rewards play a key role. About the transition from active to dormant (or, recovered, $R$) state, getting bored, divergence of attention and forgetting are the major contributors (in the transition rate, $\gamma$, Fig. \ref{fig:model_dynamics}(a)). Above all, in a key finding, we see that there is enough room for people to switch back from dormant to active state, and become broadcasters again; while the major reasons behind this are reminders by friends (for the forgetful customers) and substantial win by a friend (for the bored customers), \textit{remarketing}, or reminders by firms, also can contribute to some extent in this transition. Table 1 summarizes our results, with the physical interpretation of the parameters or rate constants of each of these above mentioned transitions.

\begin{table}[tbp]
\centering
\begin{tabular}{c}
\includegraphics[width=9cm]{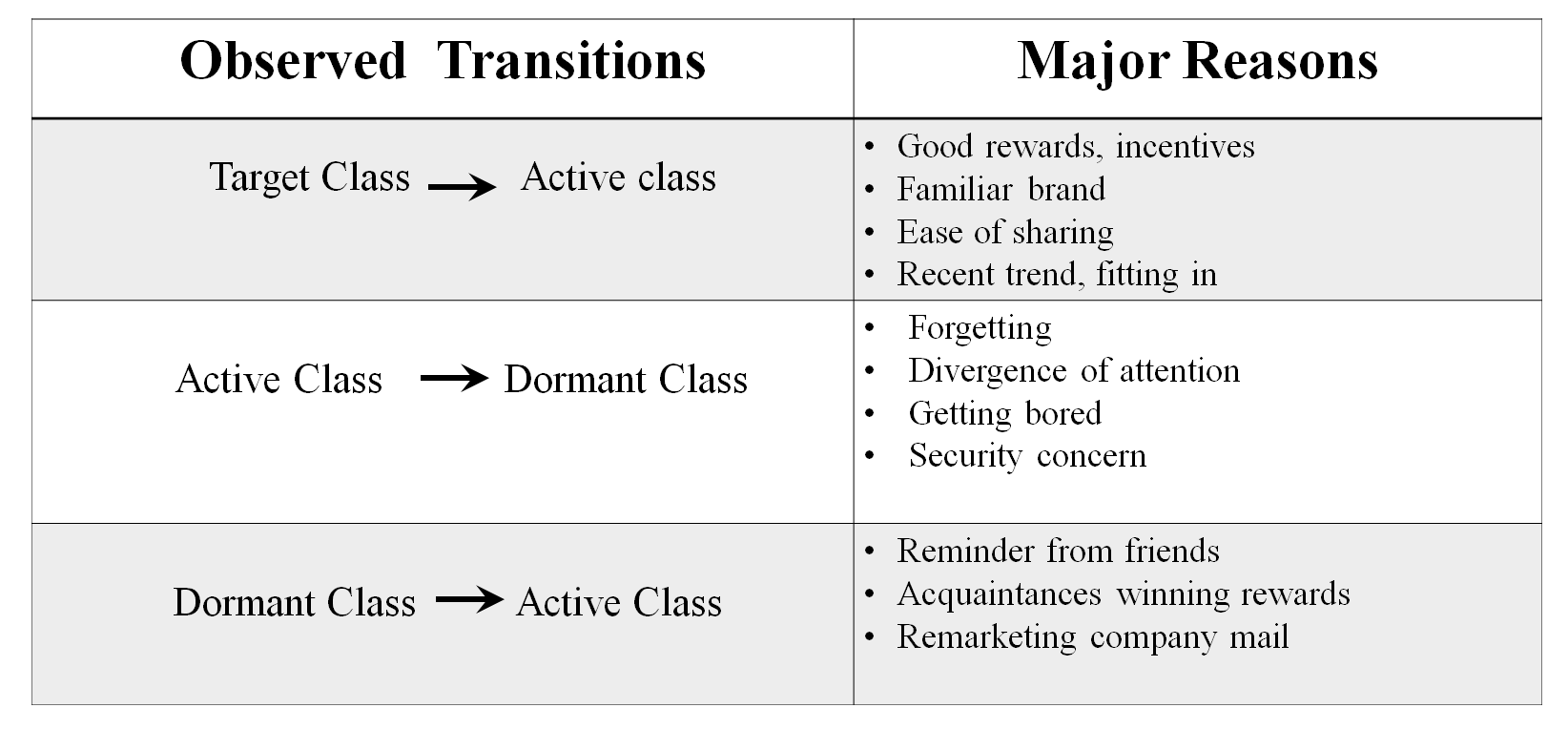}
\end{tabular}
\caption{Physical interpretation of the major parameters that contribute in switching}
\end{table}
\section{Discussions and Conclusion}
From the survey we see that the dynamics of VM propagation is a complicated nonlinear phenomenon. It involves several interactions between the participants, causing them to switch from one subpopulation to other. Influenced by several intensive and extensive parameters, the participants can affect the sharing decisions of their neighbors as well. Here, we wish to point out some managerial implications of our results. Companies should design their strategies to boost the behavioral switching of the customers in their own favor by keeping some key factors like reward, security, ease of share etc. in mind. Influencing the customers who are lost due to reasons like forgetting or attention diversion to rejoin the campaign could be a very important aspect of a successful campaign. A calculated and measured reminder sent from the firms' end without bothering the customer much, could also be a strategy to gain the lost customers back. \\
We also observe that people are more likely to go through a transition back to the active class, which is beneficial for the company, if the recommendation comes from a family member or a trusted friend, instead of the company. We conclude that these recommendations intrinsically act as an authentication of the VM campaign. We extrapolate that the firms need to figure out an indirect way of reminding customers through their friends. One possible way could be using discount coupons, where a satisfied customer of a product can share one coupon with a friend or family member who is not currently using the product. Coupons with intelligently scripted content can provide multidimensional positive benefits to the coupon sharer, the coupon receiver, and the firm. Recent studies show that if modified accordingly, this can also be used as a strategy to enhance customers' social empowerment \cite{hanson2018friends}. This could be one way to increase the close neighbor interaction with the dormant class.  \\
The goal of our study was to understand and design a conceptual framework for referral marketing campaigns that go viral on social media through an extensive survey. Analyzing the data collected from 331 participants, we successfully pointed out several key guiding factors of the dynamics,  and identified their implications in a mathematical ODE epidemiology model. In future, we intend to pursue a more detailed study based on survey results to figure out, correlations, if any, between these multiple elements we identified which shape the dynamics of a VM campaign. Moreover, we plan to introduce more realistic interactions on the top of the simple SIR dynamics, to depict a real world scenario. The model will form the basic structure of VM dynamics incorporating some important inputs like inherent forgetfulness of the customers and possible reminders.

\bibliographystyle{elsstyle}
\bibliography{references}
\newpage
\appendix
\section{Questionnaire}\label{sec:appA}
\textbf{Viral Marketing  or viral advertising is a business strategy that uses existing social networks to promote a product. Generally, customers get rewards/ offers for this sharing.  For example you can imagine the rewards given by Google Tez, or storage space given by Dropbox when you recommend the product(s) to your friends.  All the questionnaires are related to offers and rewards that require online sharing. }
\newline
\begin{center}
\textbf{Module 1}
\end{center}

\begin{enumerate}
\item Age: \\

\rule{3cm}{0.001mm}
\item Gender:  \\

O Male \\
O Female \\
O Other\\
\item Have you ever received an email/message from a friend about a product that gives offer/reward if you forward the message to your other friends?  \\

O Yes\\
O No \\

\item To avail an offer/reward/discount, do you share an online advertisement with a friend which you received from your social network profile(s)? \\

O Yes\\
O No \\
\end{enumerate}

\begin{center}
\vspace{5cm}\textbf{Module 2}
\end{center}

\begin{enumerate}
\item What made you not to share the viral offer you received?\\

O Not felt secured \\
O The process was long and the profit was low \\
O Not sure about hidden clauses\\
O Other \rule{7cm}{0.001mm} \\
\item Will reading or hearing a positive review from the company website/company's promotional emails about a viral offer change your decision about availing it?  \\

O Yes\\
O No \\
O Maybe \\
\item If you receive a particular offer message from multiple friends, OR that brand/offer is frequently mentioned in your social circle, will you be interested to try the offer? \\

O Yes\\
O No \\
O Maybe \\
\item Is there any condition (Other Than Above) in which you will be interested to try a viral offer?\\
O Yes\\
O No \\
\item If `yes', please mention some conditions (For example, 'heard my friend/ several others are getting benefit from the offer' etc.)\\

 \rule{15cm}{0.001mm} \\
\end{enumerate}

\begin{center}
\vspace{4cm}\textbf{Module 3}
\end{center}

\begin{enumerate}
\item What are the criteria that you check while availing a viral offer? (You may select multiple options)? \\

O Product brand \\
O Amount of reward you get for sharing\\
O Easy to share \\
O How many friends recommended the product to you\\
O Recent trends\\
\item Have you ever missed a viral offer that you decided to avail?\\

O Yes\\
O No \\
\item If `yes', then please specify the reason(s) \\

O I forgot about it\\
O I lost the message \\
O My attention got diverted \\
O Other \rule{4cm}{0.001mm} \\
\item Have you ever missed /forgot about /decided not to avail a viral offer, and then finally took it later, after your friend/ family member reminded you OR gave you a good review? \\

O Yes\\
O No \\
\item If `yes', please specify a case (For example: 'Did not know all the advantages of the offer earlier', 'got several recommendations later' etc. )\\
\\
\rule{15cm}{0.001mm} \\
\item Do you forward viral offers sent to you earlier before availing a service to get some benefits? (for example, before suddenly booking movie tickets or flight tickets, will you check for a message which can give you an extra discount, if you forward it to five of your friends?) \\

O Yes\\
O No \\
\item Have you ever lost interest in an ongoing viral offer?\\

O Yes\\
O No \\
\item If `yes', then specify the reason(s)?\\
O Profit was not good compared to the efforts\\
O Received better offer\\
O Not felt secured \\
O Got bored \\
O Other \rule{4cm}{0.005cm} \\
\item Will you be again interested in an offer that you decided to ignore, if you hear some of your friends have won good rewards from it?\\

O Yes\\
O No \\
O Maybe\\
\end{enumerate}
\section{ Cover Letter}\label{sec:appB}
\noindent Dear all, \\
We are currently working on the effects of friends on someone's online marketing habits. In order to get some insight about online marketing, we would appreciate it if you can  take the time to fill in a very short questionnaire. The survey should take around 2-3 minutes to complete.\\ \\
PLEASE REMEMBER that the survey is about the offers that ask you to forward the messages to your friends before availing the offer. This marketing strategy is known as 'viral marketing' and all the questions in this survey are associated with this particular marketing strategy.
\\ \\
Please go to the link:    \url{https://goo.gl/f4oxXV}
\\ \\
The survey is completely anonymous and uses Google form. The answers you provide will be kept confidential. It is to notify you that by responding thorough the link associated with this email, you are giving your conscent for the use of your response in our research purpose only. If you have any problem in accessing this questionnaire or are unsure of any question please email XXXXX@iitk.ac.in.\\
Thank you for taking the time to read this email.\\ \\
Regards\\
XXXX XXXXXX

\end{document}